\documentclass[twocolumn]{elsart}

\usepackage{graphics}
\usepackage{color}
\usepackage{amssymb}
\usepackage{amsmath}
\usepackage{pifont}
\usepackage{lscape}

\begin{document}

\begin{frontmatter}

\title{Extraction of $N^*$ information from the limited
$p(\gamma,K^+)\Lambda$ data set}

\author[gent]{S. Janssen},
\ead{stijn.janssen@rug.ac.be} 
\author[glasgow]{D. G. Ireland}, and
\ead{d.ireland@physics.gla.ac.uk} 
\author[gent]{J. Ryckebusch}
\ead{jan.ryckebusch@rug.ac.be} 

\address[gent]{Department of Subatomic and Radiation Physics, \\ 
Ghent University, Proeftuinstraat 86, B-9000 Gent, Belgium}

\address[glasgow]{Department of Physics \& Astronomy, \\
University of Glasgow, Glasgow, G12 8QQ, Scotland, UK}

\begin{abstract}

Nucleon resonance information is often obtained from fitting
hadrodynamical model calculations to data, where model parameters such
as resonance coupling constants are the free parameters in the fitting
procedure. For reactions with a limited data set, such as $p(\gamma,
K^+)\Lambda$, complications in the extraction of reliable $N^*$
information occur not only through theoretical uncertainties, but also
due to technical difficulties in the fitting procedure. In this
article we outline a fitting strategy based on a genetic algorithm and
illustrate the kind of ambiguities which can arise.

\end{abstract}


\end{frontmatter}

Understanding the composite, quark-gluon structure of hadrons is a
major topic in intermediate energy nuclear physics. One way to
investigate this system is to study its excitation spectrum, as it is
a direct consequence of the hadron's composite character. At present,
the experimental identification of the full excitation spectrum is far
from complete, and most of our knowledge is based on analyses of pion
induced or pion production reactions.

A complete understanding of the nucleon resonances can
only be achieved by a full, coupled-channel analysis which
incorporates reactions involving all possible initial and final states
to which nucleon resonances can couple
\cite{Manley,Vrana,Penner}. Meson production reactions such as 
$\gamma N \rightarrow \pi N$, $\pi \pi N$, $\eta N$, $\omega N$, $K
\Lambda$, $K \Sigma$, $\ldots$, should be included, as well as Compton
scattering and meson-induced production reactions. It is clear that a
simultaneous description of all such reactions is an enormous task.

When the (photoinduced) meson production processes are treated at a
hadronic level, the resonances are described by effective fields with
independent quantities such as mass, strong decay widths and photo
helicity amplitudes. Usually one determines these resonance parameters
by fitting model calculations to the available data. The fitting
procedures are therefore a necessary step in the extraction of
resonance information from experimental measurements. However, by
taking into account all known resonances in a full coupled-channel
calculation, one would require a few {\em hundred} parameters to be
determined. As a first step, a single-channel analysis of one
particular reaction is essential to identify general features, and to
simplify the problem by reducing it to one dependent on only a few
{\em tens} of parameters.

The search for an optimum fit to the data, by a model depending on
what is still a large number of parameters, is not
straightforward. The situation is considerably worsened when reaction
channels such as $\omega N$, $K \Lambda$ or $K \Sigma$ are examined,
since only limited data sets are available.  In these cases, one may
encounter the situation where model parameters are not sufficiently
constrained by the restricted data sets and similar qualities-of-fit
are obtained with different sets of parameter values. In this paper,
we highlight the issue by presenting a fitting procedure based on a
genetic algorithm (GA) and commenting on some of the related
ambiguities which arise in the complex fitting processes.  We have
chosen the $p(\gamma,K^+)\Lambda$ reaction as an example with a
limited amount of published data, and where we can restrict the number
of free parameters by employing a single-channel model.

In our model, the $p(\gamma,K^+)\Lambda$ reaction dynamics is
described in terms of hadronic degrees of freedom by adopting an
effective Lagrangian approach \cite{Janssen_backgr}. The tree level
Feynman diagrams contain the usual Born terms, the $K^*(892)$ and
$K_1(1270)$ $t$-channel mesons and two hyperon resonances
$S_{01}(1800)$ and $P_{01}(1810)$ in the $u$-channel. All these terms
constitute the so-called {\em background}. In the $s$-channel, the
nucleon resonances $S_{11}(1650)$, $P_{11}(1710)$, $P_{13}(1720)$ and
$D_{13}(1895)$ are included. Note that the $D_{13}(1895)$ is a
candidate for a ``missing'' nucleon resonance \cite{Mart2}. The finite
extension of the meson-baryon vertices is implemented by the use of
hadronic form factors \cite{Haberzettl_gauge,Davidson}. In
Ref.~\cite{Janssen_backgr} we stressed the difficulties associated
with parameterizing the background diagrams in
${p(\gamma,K^+)\Lambda}$ calculations and presented results for three
plausible background schemes. Subsequent work \cite{Janssen_elec}
showed that the ${p(e,e'K^+)\Lambda}$ process is highly selective with
respect to viable choices for dealing with the background diagrams.
The model used here is the only one which we found to reproduce
simultaneously the ${p(\gamma,K^+)\Lambda}$ and ${p(e,e'K^+)\Lambda}$
data.

All the extracted resonance parameters $G[N^*]$ are a product of an
electromagnetic and a strong coupling. A description of the various
types of resonance parameters $G[N^*]$, their normalization and their
connection to the Lagrangian structure is given in
Ref.~\cite{Janssen_backgr}. To minimize the number of free parameters,
one overall hadronic form factor cutoff mass for all Born terms and
one for all resonance diagrams are introduced. Those coupling
constants and cutoff masses constitute the free parameters of the
model. In the present hadrodynamical model, this amounts to 22 free
parameters which have to be determined by the data. We have used the
SAPHIR data set \cite{Tran} which contain 24 total cross section, 90
differential cross section and 12 recoil polarization asymmetry
points, in the photon energy range from threshold up to 2 GeV.  The
fits are performed by minimizing $\chi^2$. Finding the global $\chi^2$
minimum in a 22-dimensional parameter space is not trivial since
although some of the parameters are constrained to a specific range by
physics arguments, in principle they are all unknown quantities.

In order to minimize the $\chi^2$ function and extract $N^*$
information from the $p(\gamma,K^+)\Lambda$ data, we have used a {\em
genetic algorithm}. In this strategy, a set of solutions is randomly
generated. Each solution $\mathbf{p}$ is an encoding of trial values
of the free parameters, and is used to evaluate the $\chi^2 \left(
\mathbf{p} \right)$ function which determines its ``fitness''. The
population is then {}``evolved'' in a manner analogous to biological
evolution, either by mixing values (``crossover'') or by changing
values at random (``mutation'').  The population gradually migrates to
one or more better points in parameter space. Whilst these optima are
not guaranteed to be global, GA research \cite{gol89,dav91} has shown
that {}``reasonable'' solutions can be found very efficiently.

The output of the GA is the individual with the best fitness after a
prescribed number of fitness function evaluations. The parameters
associated with this individual are then used as input to the Minuit
\cite{min95} package which employs a variation of the
Davidon-Fletcher-Powell variable-metric minimization algorithm. In
this way we take advantage of the wide search capabilities of the GA,
but use the strength of a traditional optimizer to obtain final
solutions. 

\begin{figure}
\centering
\resizebox{0.45\textwidth}{!}{\includegraphics{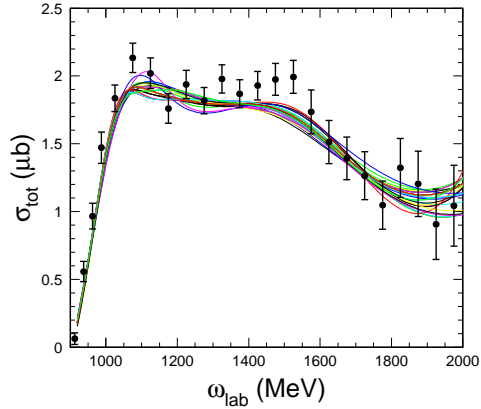}}
\caption{\label{fig:totcs}Model calculations for the energy
distribution of the total $p(\gamma,K^+)\Lambda$ cross section. The
curves correspond to the 38 solutions with $\chi ^2 \le 3$. The data
are from Ref.~\protect\cite{Tran}.}
\end{figure}

\begin{figure}
\centering
\resizebox{0.48\textwidth}{!}{\includegraphics{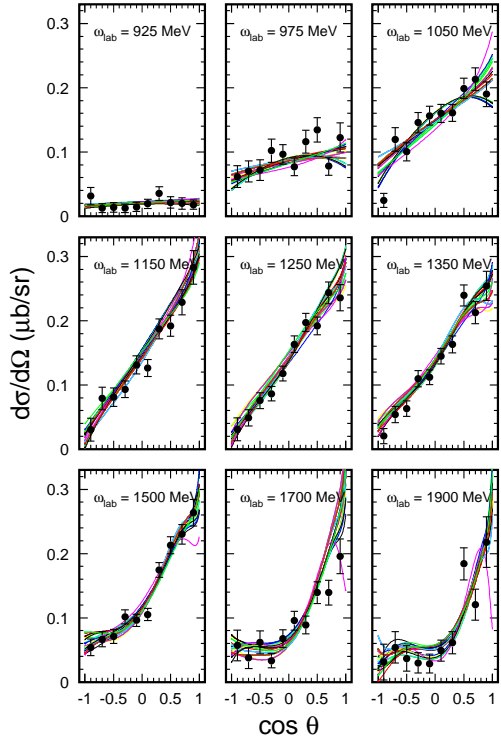}}
\caption{\label{fig:diffcs}Model calculations as in
  Fig.~\ref{fig:totcs} but for the angular distribution of the
  differential $p(\gamma,K^+)\Lambda$ cross section. The data are from
  Ref.~\protect\cite{Tran}.}
\end{figure}

\begin{figure}
\centering
\resizebox{0.45\textwidth}{!}{\includegraphics{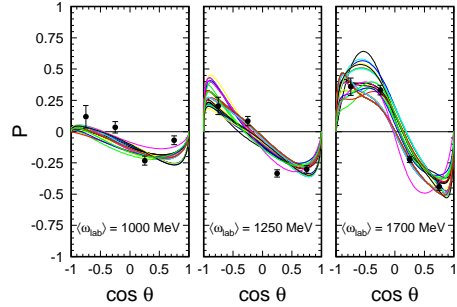}}
\caption{\label{fig:recpol}Model calculations as in
  Fig.~\ref{fig:totcs} but for the angular distribution of the
  $p(\gamma,K^+)\vec{\Lambda}$ recoil polarization asymmetry. The data
  are from Ref.~\protect\cite{Tran}.}
\end{figure}

To test whether global optima were being found, 50 GA plus Minuit
calculations (run concurrently on a Linux-based PC farm) were carried
out. Of these calculations, 38 resulted in Minuit converging with a
$\chi^2$ below 3.00. This was thought to be a reasonable value, since
our previous work \cite{Janssen_backgr} had
shown $\chi^2$ values of just under 3.00 to be achievable. The range
of the fitness of these solutions\footnote{From now on, we refer to
``solutions'' as meaning those in the converged set of 38.} was 2.45 $
\leq \chi^2 \leq$ 2.93, indicating a reasonable goodness-of-fit to the
available data for {\it each} solution. This is apparent in
Figs.~\ref{fig:totcs}, \ref{fig:diffcs}, and \ref{fig:recpol} where
the model calculations for the total and differential cross section,
and the recoil polarization asymmetry are plotted against SAPHIR data.

\begin{figure*}
\centering
\resizebox{0.95\textwidth}{!}{\includegraphics{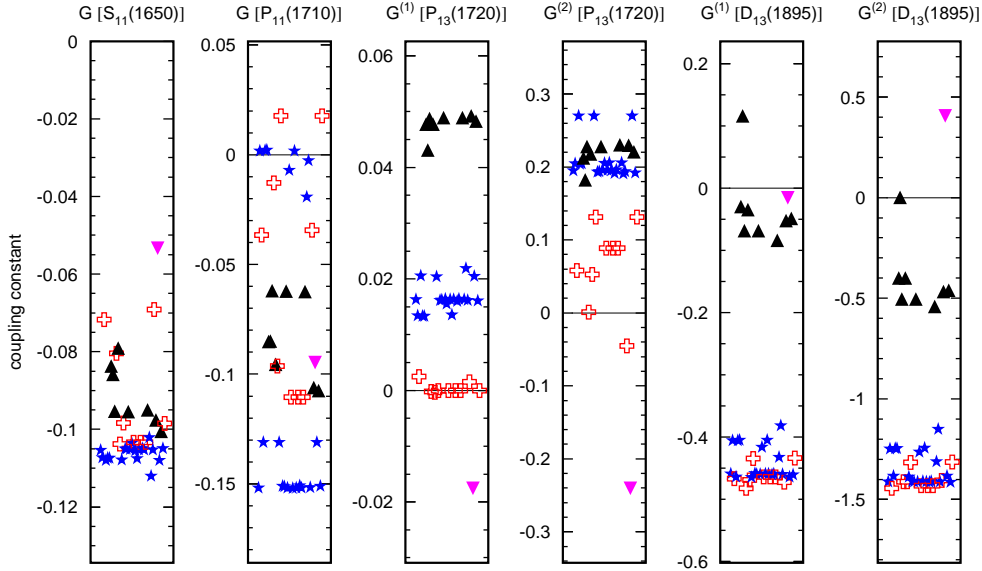}}
\caption{\label{fig:stab}Values of the coupling constants of the four
  $N^*$ resonances $S_{11}(1650)$, $P_{11}(1710)$, $P_{13}(1720)$ and
  $D_{13}(1895)$ in the reaction dynamics of
  $p(\gamma,K^+)\Lambda$. They correspond to the best solutions of the
  minimization procedure. The four clusters are labeled by \ding{115}
  \ding{116}, \ding{072} and \ding{57}.}
\end{figure*}

Given this similarity in the description of the available data, it is
interesting to investigate how close these solutions of the
minimization procedure are to each other in parameter space. In
Fig.~\ref{fig:stab}, resonance parameters for all the solutions are
plotted. At a first glance, it appears that the resulting $\chi^2$
surface in the free parameter space is extremely erratic and has a
large number of local minima. The solutions for $G[P_{11}(1710)]$ show
that the value could be anything within a wide range. On the other
hand, the $S_{11}(1650)$ coupling constant exhibits clustering in one
region, indicating that the value must be within a reasonably small
range. Of great interest is the clustering exhibited in the
$G^{\left(1 \right)}\left[P_{13}\left(1720 \right) \right]$ coupling
constant. Here, there are three distinct clusters plus one isolated
point. Furthermore, two clusters appear in the $D_{13}$ parameters. We
are therefore unable to pin down clear-cut values for all $N^*$
parameters. This clearly illustrates the shortcomings in extracting
reliable resonance parameters from the available limited
$p(\gamma,K^+)\Lambda$ data.

To study how the coupling constants are correlated, we have assigned
each of the four clusters in the $G^{\left(1
\right)}\left[P_{13}\left(1720 \right) \right]$ parameter a distinct
symbol. In Fig.~\ref{fig:stab}, there appears to be a significant
correlation between the $P_{13}$ parameter and the $D_{13}$
couplings. In the $D_{13}$, the \ding{115} and \ding{116} clusters
form one subset, whilst the \ding{072} and \ding{57} form
another. Since the role of the $D_{13}(1895)$ in this
$p(\gamma,K^+)\Lambda$ reaction is still under debate, we have split
the solutions into two subsets according to this clustering in the
$D_{13}$ parameters. The characteristics of the two subsets are
summarized in Table~\ref{tab:chi_values}.

\begin{table}
\begin{center}
\begin{tabular}{ccc}
\hline \hline
Subset & 1 & 2 \\
\hline
$\langle \chi^2  \rangle $ & 2.56 & 2.69 \\
$\sigma_{\chi^2}$ & 0.13 & 0.12\\
best $\chi^2$ & 2.45 & 2.50 \\
number of points & 9 & 29 \\
cluster labels & \ding{115} \ding{116} & \ding{072} \ding{57} \\
\hline \hline
\end{tabular}
\caption{Characteristics of the two subsets localized in the solutions
  of the minimization procedure. The averaged $\langle \chi^2
  \rangle$, its standard deviation and the best $\chi^2$ are given
  together with the number of points in the set and the corresponding
  labels in Fig~\ref{fig:stab}.}
\label{tab:chi_values}
\end{center}
\end{table}

\begin{figure}
\centering
\resizebox{0.43\textwidth}{!}{\includegraphics{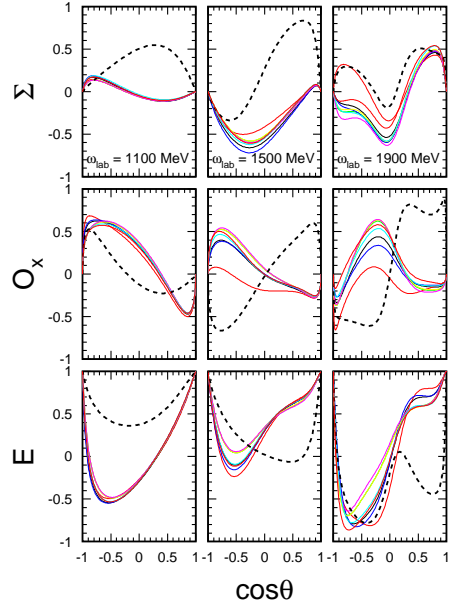}}
\caption{\label{fig:asyms.cl1} Model predictions with the solutions of
subset 1 (\ding{115} and \ding{116}) for the angular distribution at
various photon energies of the $p(\gamma,K^+)\Lambda$ photon beam
asymmetry ($\Sigma$), the beam-recoil asymmetry ($O_x$) and the
beam-target symmetry ($E$).}
\end{figure}

\begin{figure}
\centering
\resizebox{0.43\textwidth}{!}{\includegraphics{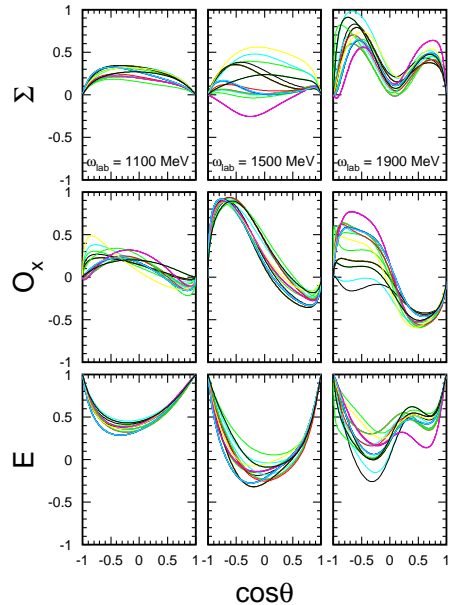}}
\caption{\label{fig:asyms.cl23} Model calculations as in
Fig.~\ref{fig:asyms.cl1} but with the subset 2 (\ding{72} and
\ding{57}).}
\end{figure}

As a further study, we show model predictions for the two subsets for
some unmeasured observables. In Figs.~\ref{fig:asyms.cl1} and
\ref{fig:asyms.cl23} this is done for the photon beam asymmetry
$\Sigma$, the beam-recoil asymmetry $O_x$ and the beam-target
asymmetry $E$. Since none of these asymmetries for the
$p(\gamma,K^+)\Lambda$ reaction are available yet, they do not
represent constraints in the minimization procedure.  Within each of
the two subsets, the predicted kaon-angle dependence of $\Sigma$,
$O_x$ and $E$ can be seen to follow similar trends (the dashed
exception in Fig.~\ref{fig:asyms.cl1} is the solitary \ding{116}
solution). 

More important discrepancies are observed when comparing the asymmetry
distributions of the two subsets. The main difference appears to be a
result of the value of the $D_{13}(1895)$ resonance parameters.  It is
interesting to note that the value of the $G\left[ P_{11} \left(1710
\right) \right]$ parameter appears to have only a small influence in the
asymmetries.  Reliable extraction of $P_{11}\left(1710 \right)$
information from $p(\gamma,K^+)\Lambda$ data is therefore likely to be
difficult, even when more (double) polarization asymmetries are
available. The model predictions in Fig.~\ref{fig:asyms.cl23} show
that also the differences in the $P_{13}$ parameters between the
\ding{72} and \ding{57} solutions do not translate to significant
differences in the plotted asymmetries. On the other hand, the
calculation with the ``odd'' behavior in Fig.~\ref{fig:asyms.cl1} is
the solitary \ding{116} solution. Since this solution deviates from
the \ding{115} cluster most strongly in the $P_{13}$ parameters, this
shows a non-negligible influence of the $P_{13}(1720)$ resonance on
the observables.


The calculations show that the present data can not discriminate
between two main regions in parameters space. The clear differences in
the asymmetries between the two subsets will require only modest
statistical accuracy to distinguish experimentally. Therefore, the
upcoming data for the photon beam asymmetry \cite{Zegers,DAngelo} and
the double polarized asymmetries \cite{Klein} will be vital in
constraining the values of resonance coupling constants and
understanding the reaction mechanism underlying $K^+ \Lambda$
production.

In conclusion, we have developed a fitting strategy based on a genetic
algorithm in order to extract resonance information from
$p(\gamma,K^+)\Lambda$ data. The advantages of the genetic algorithm
are that the available parameter space is extensively sampled, and
that the large number of calculations allows one to investigate any
ambiguities in the solutions.  The success of this strategy perhaps
points to its use in problems with similar or larger numbers of free
parameters.

The solutions obtained from the minimization runs result in model
calculations in close agreement with the available data set. However,
from a detailed exploration of the parameter values, it appears that
not all solutions are contained in the same region of parameter
space. For some of the variables, large deviations appear and no well
defined value can be pinned down. This is particularly true for the
$P_{11} \left(1710 \right)$ coupling constant.  For other $N^*$
parameters, clustering in a specific area is observed. One stable
cluster appears for the $S_{11}\left(1650 \right)$ parameter and for
the $P_{13} \left(1720 \right)$ and $D_{13} \left(1895 \right)$, a few
isolated groups of solutions are identified. On the basis of these
clusters in the $N^*$ parameters, we were able to select two different
subsets of solutions of the $\chi^2$ minimization. Solutions within
each subset produce more or less identical predictions for unmeasured
polarization observables, but large differences were observed when
comparing the two sets. 

The calculations indicate that the extraction of reliable resonance
information from a limited data set is by no means trivial, and has to
be handled with the greatest of care. An important conclusion is that
when using theoretical models to analyze the excitation spectrum of
the nucleon based on a limited data set, the accuracy of predictions
will be greatly affected by uncertainties stemming from the fitting
process. Hence, employing a strategy which enables a thorough
exploration of parameter space is essential. We have shown that the
upcoming measurements of specific polarization asymmetries are
essential to reduce the uncertainties resulting from the present
limited data set.

{\bf Acknowledgments} S.J. wants to thank the University of Glasgow,
where part of this work was completed, for the hospitality during his
stays.

\bibliographystyle{elsart-num}

\end{document}